# *Attilio Sacripanti*

*Judo how to enhance tactics in competition: biomechanics of combination and action reaction attacks.*

**I Introduction**

**II Initiative in high level competition**

**III The Classical Japanese way**

III.1 ***Sen*** [ *Renzoku and Renraku Waza*]

III.2 ***Go no Sen*** [*Bogyo Waza*]

III.3 ***Sen no Sen*** [*Kaeshi Waza*]

**IV Biomechanics of Initiative**

**V Initiative as exploitation of the kinetic energy and angular momentum**

**VI Current application**

VI.1 ***Direct Attack***

VI.2 ***Combination***

VI.2.1 *Combination Examples from high level competitions*

VI.3 ***Action Reaction***

VI.3.1 *Action Reaction Examples from high level competitions*

**VII Conclusions**

**VIII References**

*Judo how to enhance tactics in competition: biomechanics of combination and action reaction attacks.*

by *Attilio Sacripanti*\*†‡§\*\*

\*ENEA (National Agency for Environment Technological Innovation and Energy) Robotic Laboratory
†University of Rome II "Tor Vergata", Italy
‡FIJLKAM Italian Judo Wrestling and Karate Federation
§European Judo Union Knowledge Commission Commissioner
\*\*European Judo Union Education Commission Scientific Consultant

## Abstract

This paper is an ideal continuation of the previous one "**How to enhance effectiveness of Direct Attack Judo throws** " in it there are analyzed the two following parts of Judo tactics in high level competitions: combination, and action-reaction.

The paper start with a review of the Japanese approach to Initiative ( Sen) and follows by the biomechanical view of the same subject.

High level competitions are the main argument not only of coaches match analysis , but also of a lot of scientific researches.

However the connection between these two field that analyze the same subject is very hard . A lot of information are not easily transfer to coaching area.

In this paper the effort to give coaching useful information is the primary aspect also at detriment of some formal mechanical information.

After a new Operative Classification of throwing techniques , the biomechanical analysis of combination and action-reaction tricks flows in easy way singling out some interesting finding, useful for coaches. With this effort the biomechanical analysis of judo Interaction in high level competition could be considered completed.

*All technical sequences photographs, thanks to David Finch*

# Attilio Sacripanti

# Judo how to enhance tactics in competition: biomechanics of combination and action reaction attacks

## I Introduction

In modern time, high level competitions are characterized by a dynamic approach to the fight, but if we analyze the athletes' dynamic behavior deeper, it is possible to find some important remarks useful for understanding the most difficult kernel of competitions: tactics.
If we glance at the Couple of Athletes system, in judo competition this Couple moves with very quiet pace, to study and analyze the adversaries' behavior, finding a weak spot both in grips or body's movements.
If this spot is found then lightning attacks start, to seize on this moment and to grab a technical vantage till to victory (Ippon). On the basis of this description it seems obvious that fast dynamics events only during a throwing action, called in Biomechanical terms *Interaction* between Athletes.
In this paper we will try to describe in Biomechanics way the most important situations classified by match analysis that arise in high level competitions, finding the common background in order to let easier to enhance, from the teaching or coaching point of view, these tactical situations that occur more often in judo competitions. Quantitatively speaking the shifting velocity of couple range between ($0.2 \leq v \leq 0.4$ m/s) obtaining the mean distance covered by judokas ( 121,1 m) .[1].
Otherwise the mean attack speed range between 1.3 and 1,8 m/s more or less five times the shifting velocity.

## II Initiative in high level competition

In modern time the best way to gain vantage during judo competitions is always to grab the initiative in fighting. But what means in plain words, to grab the initiative? Probably the best explication is: to give the fighting actions its strategy, so as to force the opponent on defensive attitude.
How it is possible to do that, by means of continuous true technical attacks that are able to put under pressure the adversary both from the psychological and physical point of view.
To apply such fighting strategy, athletes need high level of endurance, because the average energy consumption evaluated by De Goutte et alt. is indicative only of normal fighting strategy [2].
They show that a normal judo match induces both protein and lipid metabolism even if the anaerobic system is brought into action, with mean levels of plasma lactate of 12.3 mmol/l., if we remember normally Couple shifting velocity range between 0,2-0,4 m/s, with attack speed 4 or 5 times faster, that depends by the kind of technique applied.
Then in terms of fighting strategy Athlete must grabs the initiative in the way previously proposed, to achieve a sensible psychological vantage that can be changed in increased probability to win only by means of tactics. In other words, athletes can grab initiative by means of strategy of repeated attack actions and solve in winning way the fight by means of tactics.
Then Strategy is defined as the plan or the flexible connection of more plans based on the coordination of physical efforts, harmonized with relative movement finalized to fights' victory, instead Tactics is defined as the capability to utilize in right way the transitory phase for victory. On the basis of these definitions it is possible to understand the deep difference between these two activities.

A strategic plan can be studied and coached in advance, and then it is possible to connect it to the rational analysis of the fight. While tactical capability is essentially founded on instant intuition of technical action, then it is not possible to teach it, in any way (it is a special skill of a Champion).

At light of the previous definition, it raises a problem: if Tactics is a gift of a Champion, for non-Champion it is possible to train it in effective way? Yes, the solution like every situation sports is to repeat and solve during the technical training the "strange" situations happened in real competition, the repetition will produce the capability to solve same or lightly different situations in real competitions.

From this point of view Match analysis is a powerful and useful tool. [3]

## III  The Classical Japanese way

In Japan in the old time there were many ancient schools of martial arts that have studied initiative, until to realize "*Densho*" books about secret principles, wealth of each school.

Most interesting is the old Japanese approach the initiative during fights or competition. .

The high competition is matter of deep studies and specific analysis, with modern tools of Match Analysis, normally such studies are finalized to understand the strong and weak points of Athletes both of the own Team and adversaries (scouting and biomechanical evaluation ) [4].

In the *Densho* the way of taking the initiative is explained in three stages

In Judo fighting there are three base-form of energy utilization into action called, according to the ancient Japanese classification, *Sen, Go no sen, Sen no sen*:

**Sen**—(the initiative) needs the correct use of *Renzoku waza* (continuous applied techniques) and *Renraku waza* (consecutive techniques)

*Go no sen*—(reactive initiative) expresses by *Bogyo waza* (defensive techniques)

**Sen no sen**—(the initiative before the initiative) gets with *Kaeshi waza* (counter-attack techniques)

There is no clear distinction between *Renzoku* and *Renraku Waza* in the definition of Dictionary of Judo published by Kodokan

*Renzoku waza* ;**"the continuous application of combinations of techniques one leading into the next."**[5] pag.111.

*Renraku waza*: **"the application of several techniques in rapid succession, moving from one to the next in a smooth unbroken sequence."**[5] pag.110.

Really speaking, combinations of techniques are always several techniques connected applied in rapid succession.

However Japanese old masters and the same Judo founder Dr. Kano made a very deep analysis of such fighting actions, which for teaching reasons were grouped under the *Sen Principle* studies.

*Sen* Principle is surely the simplest and linear method to utilize the initiative by a direct and positive action. In Japan in the old time there were many ancient schools of martial arts that have studied initiative, until to realize "*Densho*" books about secret principles, wealth of each school.

To go in deep and analyze in more articulate way the Japanese approach to *sen* initiative it is possible as discriminating factor, to use time as the previous or subsequent time to apply one's special technique (*Tokui waza*), considering the time step of attack as stating point it is possible subdivide the complete attack action till to the victory, in five branches useful from the teaching point of view

- A) Direct attack with one's special (*Tokui waza*)
- B) Repeated attack with same technique in the same direction (*Renzoku Waza*)
- C) Repeated attack with continuous connected techniques *(Renzoku Waza)*
- D) Attack and variation in another technique cause of anticipated *Uke* defense (*Renraku Waza*)
- E) Attack Subsequent to a feint or action reaction attacks *(Damashi Waza)*.

## III.1 *Sen* [ *Renzoku and Renraku Waza*]

**A) Direct attack with one's special (***Tokui Waza***)**

This action is the purest application of *Sen* Principle, the decisive technique (*Kimari waza*) is directly realized, for this reason *Tori* must possess maximal kinetic energy and maximal impulse to develop maximum power in collision [6], it's fundamental that Uke opposes a weaker resistance to have success in this direct action. The attacking phase is direct, Tori needs to use his one's special *Kumi kata*, and it's also essential a less strong rival both from physical and technical point of view. This is the reason, because this kind of tactics could be ineffective against a well-prepared rival.

In the last few years, the analysis of maximum competitions [7] has proved that many athlete try to enhance effectiveness of direct attack changing the attack angle slightly during the throwing action "*to redirect into ineffective direction*" *Uke* defensive reaction.

Many other special tricks utilized in high level competitions on direct attacks to enhance their effectiveness are analyzed in a previous paper [8].

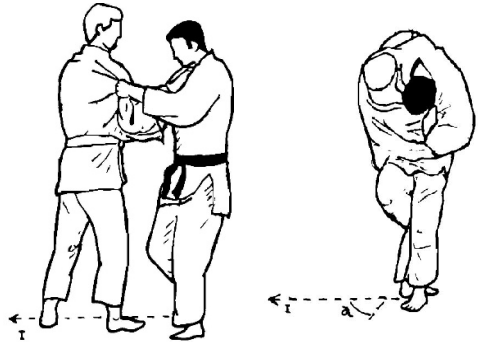

Fig 1. SEN- Direct *O Soto Gari with change in the attack angle*

Direct Match analysis performed during various judo word championships [9,10] shown the effectiveness of combinations. Japanese description on these movements during the throwing action in high dynamic situation is clearer as translated by Japanese, into the Judo Dictionary it is possible to find such specific descriptions:

*Henka suru:* **To change, to switch techniques. To shift your body to escape or evade your opponent's attack and place yourself in a position to launch an attack of your own instead, or to change from one technique of your own to another**. [5] pag. 76.

*Henka waza* :**change technique. Counter technique in which you make use of your opponent's technique to set up a technique of your own instead, or use your own techniques to set up other techniques changing quickly from one into the other.**[5] pag. 76.

*Renraku henka* : **Connection and change. For both attacking and defending, as one of the most important aspects of applying techniques effectively it is considered essential to be able create good connection between continuous applied techniques so that each evolves into the next. This can include both changing a technique of your own into another technique of your own, or changing your** *opponent's technique into a technique of your own. In either case it is necessary to cultivate your ability to move logically and efficiently from one technique into another.*[5] pag. 110

Returning to *Sen* Principle following the previous teaching classification it is possible to consider these different cases.

**B) Repeated attack with the same technique (*Renzoku Waza*)**

This kind of *Sen* Principle is applied after *Uke* stops the first attack to repeat in sequence the same *tokui waza* taking advantage or of *Uke* probable mental relax or of attack angle variation. In the first opportunity, there is the same impulse in the first and the second attack (*Nami sen*), but the second impulse will be always less or exactly the same than the one. The success of technique (*Kake* phase) and throwing (*Nage* phase) will be possible only when the second reaction is less than the previous one and contemporary Uke is more unbalanced. In the second opportunity, even if the *Uke* reaction will be the same or greater than the first one, it will be frustrated by an appropriate variation of *Tori* angle attack.

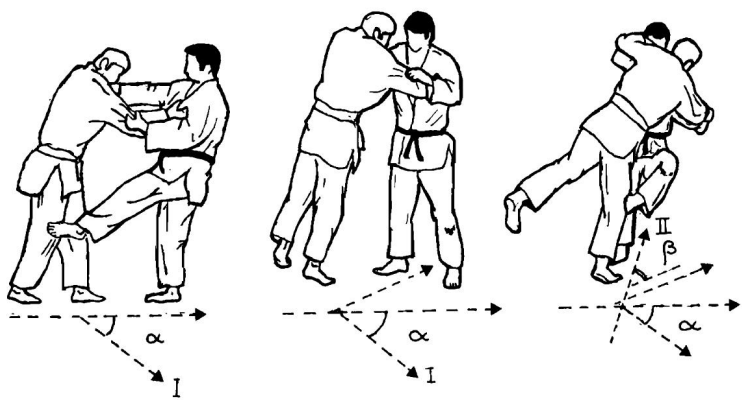

Fig 2 SEN- *Continuous Repeated throw with the same technique Hiza Guruma with angle change*

*C)* **Repeated attack with connected techniques (*Renzoku waza)***

This is the application of *Sen* Principle in connected series, using different techniques. The first action of *Tokui waza* causes *Uke* reaction who responds to a right direction to nullify the attack, but producing an opportunity of vulnerability in a second or third attack, in which *Tori,* using at best adversary's energy, applies suitably the connected technique, which is in the direction of *Uke* less resistance. The following connected techniques (*Renzoku waza*) must be considered a tactical construction, with a strategic end to change direction to the applied impulse directing it towards minimum energy trajectory (geodetic) in which there is the less resistance direction that *Uke* reaction has produced.

The analysis of the specific directions applied to two techniques combination is able to group them in three categories

- Linked techniques in the same direction (*Nami Sen)*
- Linked techniques in the opposed direction (*Gyaku Sen)*
- Linked techniques in side directions (*Yoko Sen*)

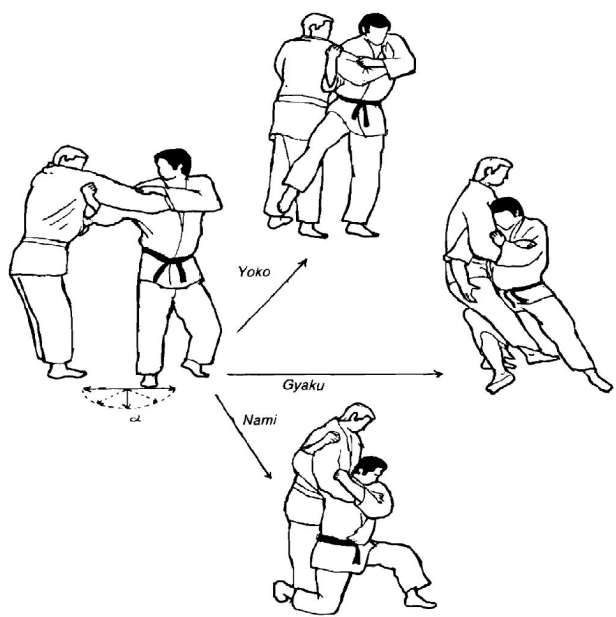

Fig 3 SEN- *Linked continuous throws on Ippon Seoi Nage*

**D) Attack and variation in another technique cause of anticipated Uke defense** *(Renraku waza)*

This is the enforcement of *Sen* Principle about the most fundamental principle "maximum efficiency with minimum effort" inside biodynamic grouping. Everything is possible if, before Tokui waza, you effect the appropriate attack trajectory variation execution, related to minimum resistance direction, cause of very advanced Uke defense.

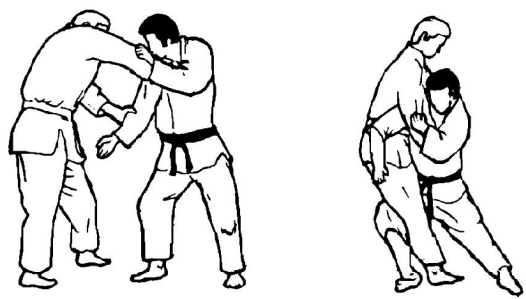

Fig 4 SEN- *Ko Uchi Makikomi attack on untimely defense on O Goshi attack*

   *F)* **Feint and subsequent attack or action reaction attack** **(*Damashi waza*)**

The *Sen* Principle application utilizing a feint; to apply the decisive throwing technique (*Kimari waza),* into Uke moving direction. It's important to realize the whole feint action in the right way allows *Tori* to keep most of his kinetic energy and to utilize it at best using *Uke* reaction.
In this way *Tori* utilizes *Uke* energy to annul the feint, adding this energy to one's own and so utilize it in his *Tokui waza* during *Kuzushi Tsukuri Kake* phase.

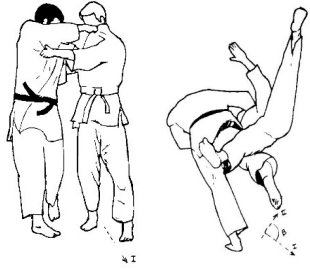

Fig.5 SEN – *Hidari Damashi Migi O Soto gari*

Noriyuki Shannoe performed an interesting evaluation of Sen by (Damashi waza) feints connecting feints to different step of throws: tsukuri and kake, and utilizing a personal technical classification (throws divided in two type with rotation and without rotation ), by binomial tests , obtained the following results if the feints was non rotational, and techniques were of the same group all the feints were connected to kake phase, in the other way if the feints were non rotational, and the throwing techniques were rotational the feints were applied during the tsukuri phase only.

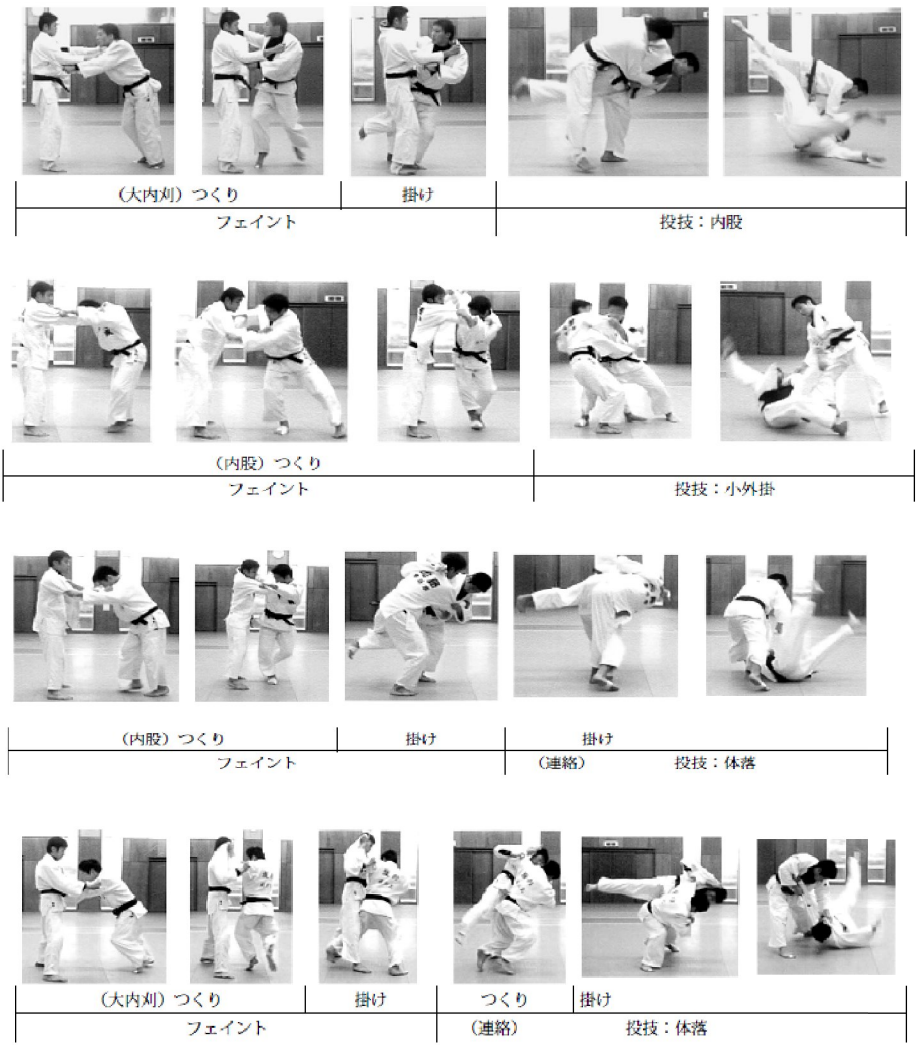

Fig 6 a,b,c,d  Shannoe experiments on Sen by (Damashi waza) connecting feints to kuzushi tsukuri.

## III.2 *Go no Sen* [*Bogyo Waza*]

The *Go no Sen* principle (the initiative contrast) is brought in some techniques called *Bogyo waza* (defense techniques).

The Japanese school, in a didactic classification, takes into consideration various possible principles about *Tori* actions which he can realize, during the attack, to oppose this attack before to effect or to develop the counter-attack.

The principles are:

A) *Go*—to break, to stop
B) *Chowa*—to avoid, to dodge
C) *Yawara*—to support, to yield
D) *Ura*—(to annul) back  [9]

It's important to note *Tori* effects his defensive technique after the ineffective *Uke* action. In general, you will find a delay time between *Uke* attack and the beginning *Tori* counter-attack. The delay time, during these defensive actions, plays a fundamental part; in fact, the counter-attack will be more efficacious and less expensive from energy point of view if, for example, it is effected at the end of *Uke* attack, cause in this way *Uke* isn't in a right force position (*null Mechanical Momentum*) and before he acquires a new balance position.

The biodynamic analysis about energetic in these actions permits to comprehend the *GO actions,* they are the most dissipative. The contrasts take place opposing to the *Uke* action the same and contrary reaction, in this way this action is completely annulled and stopped, but at the cost of considerable waste of energy; for this reason, it isn't wise, for a correct resistance direction purposes in a protracted effort, to build one's defensive action completely on these kind of technical actions.

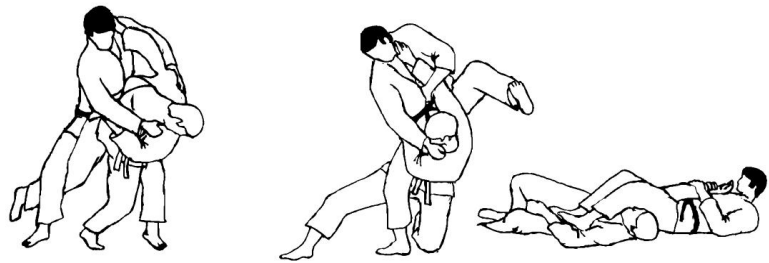

Fig 7  Go no Sen –Uchi Mata Go defense

While, *Chowa, Yawara* and *Ura* actions definitely are very close to the principle of '*maximum efficiency with minimum effort';* in fact, they are founded on a "soft" defense, these techniques utilize a lot of *Uke* kinetic energy, having a great advantage over him with minimum effort.

*Chowa* defense is essentially founded on the mobility concept (*Tai sabaki)* and on the utilization of the unbalance position, because of inefficacious *Uke* attack; it is enforced at the end of *Uke* action attack and to the less resistance direction. About *Sen* principle there are three classes of defenses related to direction:

Actions in the same attack direction (*nami chowa*)

Actions in the contrary attack direction ( *gyaku chowa*)

Actions in lateral directions to the attack (*yoko chowa*)

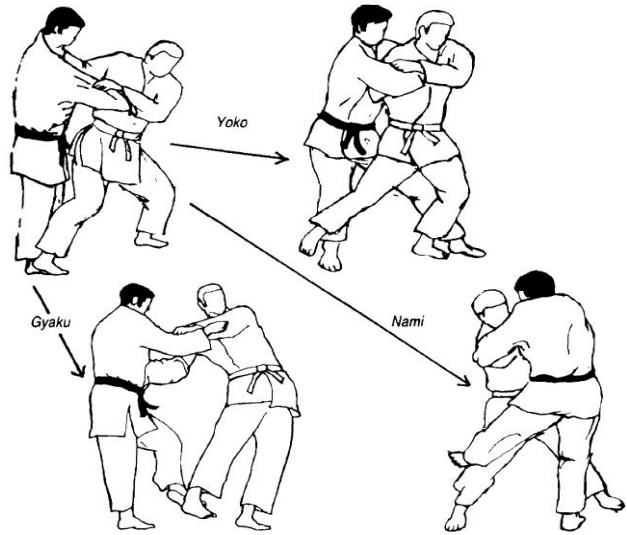

Fig 8 - Go no Sen-Tai Otoshi Chowa defence

The *Yawara* defense ( preferred by Mifune ) is even more delicate than *Chowa,* it is founded on dynamic concept of the partial kinetic energy utilization generated by *Uke* attack, this energy must be used before the attack action becomes effective and always in its same direction.

For these reasons, it's really fundamental to have considerable acrobatic capabilities and a perfect sense of timing during the execution of *yawara* defense: in fact for the principle inherent into defense, to make a mistake in timing would cause a successful of the *Uke* attack.

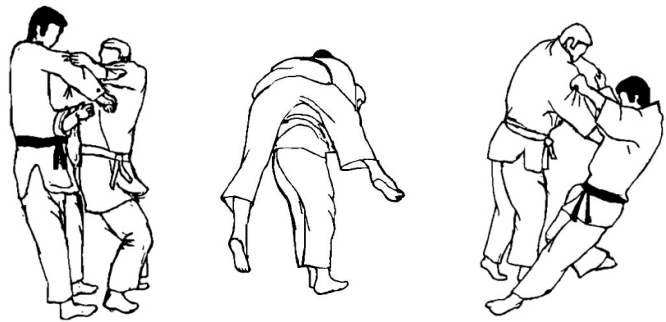

Fig 9 -Go no Sen – Morote Seoi Nage Yawara defence

The *Ura* defense is very interesting from biomechanical analysis point of view of the energy into action. In fact it is founded above all on the directional energy transformation; the physical fundamental principle of this defense is to direct most of the kinetic energy of *Uke* attack in a particular direction to effect a *nage waza* throwing. These particular techniques (*Ura nage, Te guruma, Ushiro goshi, Utsuri goshi,* etc.) are above all specific movements aim to deviate, to direct and to guide suitably the rival attack until final throwing.

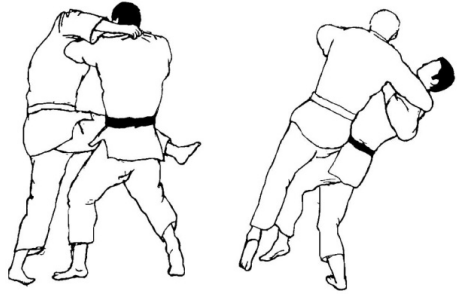

Fig 10 Go no Sen- O Soto Gari Ura Defence

It's important to remember during a competition often there are not only these forms of dynamic defense, but also "passive" defense forms founded on the exclusive use of the bio-kinetic chains and on the appropriate transfer of weight body. All these almost actions have in common the stiffening of arms to keep contact with other action in a specific situation. They are also founded on the opening grip to stop with arms the throwing of one's body. In Judo, this dangerous practice is the most frequent motive of dislocation.

## III.3  *Sen no Sen*  [*Kaeshi Waza*]

*Sen no Sen* principle (the initiative over the initiative) is perhaps the most delicate form to apply. During its execution there are two initiatives in the biodynamic grouping which mix in a single dynamic whole. The ideal action is: *Tori* realizes his counter-attack while *Uke* attack is going to start; it's an advance and superior initiative form, but it requires a very special psychophysical capability, quick reflexes and a perfect athletic form. This kind of action has the advantage to take by surprise rival during the attack keeps under his mental and physical pressure; this action has also the advantage to utilize rival's energy exploiting the initial delicate phase of the transition from potential muscular energy to kinetic energy. His practice needs a perfect and right present opportunity perception; it requires an absolute use control and a full speed in execution, to overcome *Uke* speed.

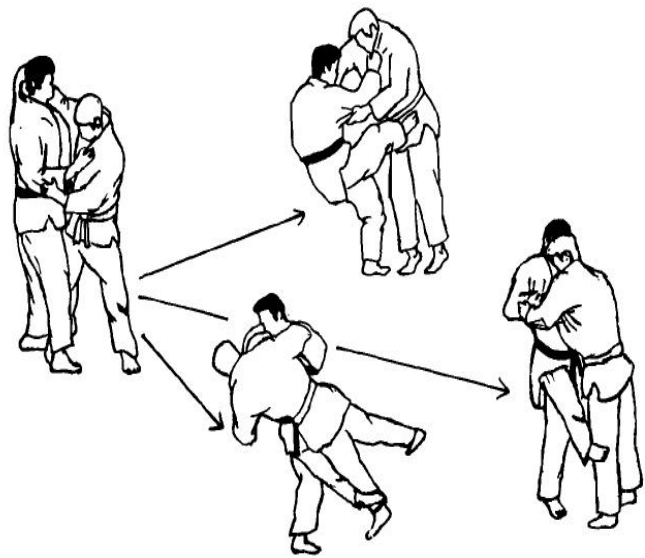

Fig 11  Sen no sen-Kaeshi against ko uchi gari attack

Contrast between classical and biomechanical vision lifting leg at different eight we have three different throwing techniques (classical vision) the same biomechanical principle application with different length of lever arm

The techniques which embody this principle are *Kaeshi waza* (counter-techniques), they are so called to emphasize fundamental aspect of opportunity during a counter-attack.

Now we'll distinguish them in this way:

Direct counter-attack in direction to attack (*Nami Kaeshi*)

Direct counter-attack in opposite direction to attack *(Gyaku Kaeshi)*

Direct counter-attack in lateral direction to attack (*Yoko Kaeshi)*

## IV  Biomechanics of Initiative

Initiative , as previously defined, is to impose to the fighting actions its strategy, so as to force the opponent on defensive attitude.

In this optic the initiative is carried out applying connected attack actions during time, this is obtained transferring to the adversary by grips or every contact point, unbalancing energy, flowing into a throws by the application of one of the two physical throwing tools ( couple or lever).

Throwing are realized by means of skilful movements that utilize not only the athletes' force but also his own body mass,  his right velocity changes and the opportune and effective body's turns (tai sabaki) .

In this optics, biomechanics could be useful to analyze and study the actions connected to the high dynamics competition that are breeding ground for the application of the Initiative.  [10]

Obviously the main thing to put under study will be the actions expressed by Tori as kinetic energy and/or angular momentum that the attacker (Tori) transfers to the receiver (Uke) applying the continuous attack strategic actions to grab the initiative.

## V  Initiative as exploitation of the kinetic energy and angular momentum

If, we're going to study the dynamic phase of competition as better utilization of owns kinetic energy, and angular momentum or the one of adversary and as problem of the dead time between preparation and attack. It's important to study the most favorable solution, from biomechanical point of view, about initiative and its better utilization.

Every biomechanical consideration, in this paper, refers to the analysis of the biodynamic grouping "athletes couple", about *Tori* actions with reference to the corresponding *Uke* actions and positions.

The study of the initiative and its utilization, leaves out of consideration concept of attack and defense; in fact the essence of defense is to grasp always the initiative, to utilize all one own and rival movements, to be able to realize a throwing technique; at this point, we can say that the better defense is an attack and vice versa.

## VI  Current studies

During these years a lot of interesting and deep papers on judo competition are wrote, and competitions have been analyzed in many different aspect and point of view.

Probably the only complaint is that the all the results are not available for coaching, for two reasons one there is not standardization on data, and second the different methodologies set up by different researchers are theoretical and useful for evaluation and not for coaching utilization.

Very few researches are produced with the goal of practical utilization.

Among these one of the most interesting is the group of researches performed in France  by Calmet and coworkers  *"Optimisation de la performance en judo : analyse des combats de haut-niveau"*  [11].

Many interesting data useful for coaching utilization are possible to find in these researches.

 In the following it is possible to see one excerpt from these works with the most interesting findings.

Fighters type  and shifting trajectories  study (Calmet, Gouriot M, 1987)

Contribution to the technical/tactical analysis of the attack by throwing techniques ( Nage waza) (Rambier R, 1987).

 Contribution to the analysis of the transition:   Tachi waza-Ne waza (P.Roux,1990).

Furthers development of the analysis transition Tachi waza –Ne waza collected by Patrick Roux Studies of attacks directions at the European Championships in 2001 and 2002.

Trainings must take into account factors related to time (structure fighting, fighting times, durations sequences, changes in technical ) ( Calmet 2006)

Kumi-kata, distances and rotations of the attacker during approaches and seizures opponent in battle unexpected attacks, techniques increasingly open and complex (Calmet 2007)

Reminders in the work of previous authors ( Calmet 2007)

In this very interesting researches complex we can find how advanced could be the collection data about the adversaries scouting.

In the following figures there shown the technical tactical organization of Marc Huizinga during the final of European Championship 2001 against Salimov

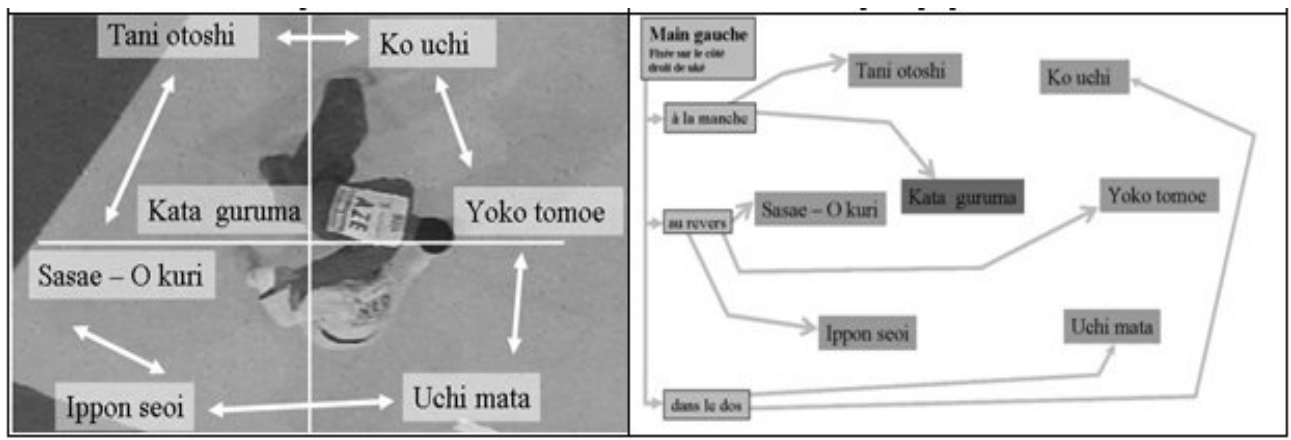

Fig 12  Technical –Tactical organization of Huizinga  against Salimov

Fig 13 Attack directions of medallists of the European Championships 2001-2002

In general among the many researches performed some others are focalized on tactics in competition or on the more frequently used techniques or in the attacks way to gain Ippon.

In the following examples are shown a series of tables collected from some of the most interesting papers that analyze high level judo competitions.

In the table 1 there is the frequency of techniques utilized by male light and heavier competitors between during the world and Olympic tournaments 1995-1999 [12]

| Techniques | ML | MH | Row Total |
|---|---|---|---|
| *Te-waza* | 575 34.35% | 183 20.09% | 758 29.32 |
| *Ashi-waza* | 590 35.24% | 365 40.07% | 955 36.94% |
| *Koshi-waza* | 74 4.42% | 64 7.03% | 138 5.34% |
| *Sutemi-waza* | 231 13.80% | 145 15.92% | 376 14.55% |
| *Osaekomi-waza* | 119 7.11% | 80 8.78% | 199 7.70% |
| *Kansetsu-waza* | 37 2.21% | 26 2.85% | 63 2.44% |
| *Shime-waza* | 30 1.79% | 6 0.66% | 36 1.39% |
| Other | 18 1.08% | 42 4.61% | 60 2.32% |
| Column Total | 1,674 64.76% | 911 35.24% | 2,585 100.00% |

Tab. 1 Frequency of techniques performed [12]

In the next table 2 are collected the decisive throwing techniques utilized also in combination at the Olympic games in Beijing 2008.

| No. | name combination throws | | | number | | % | |
|---|---|---|---|---|---|---|---|
| 1 | | | kata-guruma | 18 | | 5.94 | |
| 2 | | | morote-gari | 10 | | 3.3 | |
| 3 | | | seoi-nage | 25 | | 8.25 | |
| 4 | | | tai-otoshi | 3 | | 0.99 | |
| 5 | | | kuchiki-taoshi | 14 | | 4.62 | |
| 6 | | te waza (hand techniques) | obi-otoshi | 1 | 82 | 0.33 | 27.06 |
| 7 | | | kibisu-gaeshi | 2 | | 0.66 | |
| 8 | | | seoi-otoshi | 1 | | 0.33 | |
| 9 | | | kouchi-gaeshi | 1 | | 0.33 | |
| 10 | | | sukui-nage | 6 | | 1.98 | |
| 11 | | | sumi-otoshi | 1 | | 0.33 | |
| 12 | | koshi waza (hip techniques) | sode-tsurikomi-goshi | 6 | 9 | 1.98 | 2.97 |
| 13 | | | o goshi | 3 | | 0.99 | |
| 14 | | | uchi-mata gaeshi | 5 | | 1.65 | |
| 15 | | | uchi-mata | 13 | | 4.29 | |
| 16 | | | sasae-tsurikomi-ashi | 1 | | 0.33 | |
| 17 | nage waza (throwing techniques) | | o-uchi-gari | 9 | | 2.97 | |
| 18 | | | o-uchi-gaeshi | 1 | | 0.33 | |
| 19 | | | o-soto-gari | 7 | | 2.31 | |
| 20 | | | o-soto-gaeshi | 2 | | 0.66 | |
| 21 | | ashi waza (foot and leg techniques) | obi-tori-gaeshi | 2 | 71 | 0.66 | 23.43 |
| 22 | | | ko-uchi-gari | 10 | | 3.3 | |
| 23 | | | ko-soto-gari | 2 | | 0.66 | |
| 24 | | | ko-soto-gake | 10 | | 3.3 | |
| 25 | | | harai-goshi | 3 | | 0.99 | |
| 26 | | | hane-goshi | 1 | | 0.33 | |
| 27 | | | de-ashi-barai | 4 | | 1.32 | |
| 28 | | | ashi-guruma | 1 | | 0.33 | |
| 29 | | | rear | tomoe-nage | 10 | | 3.3 | |
| 30 | | | | sumi-gaeshi | 12 | | 3.96 | |
| 31 | | | | tani-otoshi | 5 | | 1.65 | |
| 32 | | | | yoko-otoshi | 4 | | 1.32 | |
| 33 | | | | yoko-gake | 1 | 40 | 0.33 | 13.2 |
| 34 | | sutemi waza (sacrifice techniques) | | harai-makikomi | 1 | | 0.33 | |
| 35 | | | | soto-makikomi | 3 | | 0.99 | |
| 36 | | | side | o-soto-maki-komi | 1 | | 0.33 | |
| 37 | | | | yoko-guruma | 2 | | 0.66 | |
| 38 | | | | harai-makikomi | 1 | | 0.33 | |
| total | | | | 202 | | 66.67 | |

Tab 2 most effective throws in Olympic Beijing 2008 [13]

In the next diagram there are analyzed as technical tactics the attack number applied on the right and on the left and their effectiveness during the Japan championship Open category 2003-2009 [14]

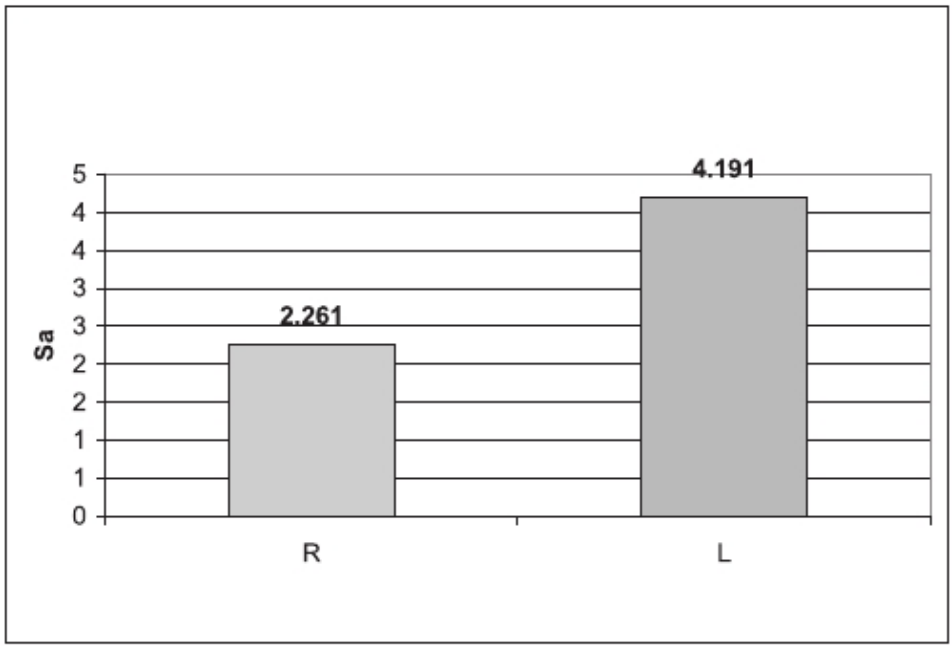

Diag 1 Side Throws efficiency during Japan Championship Open Category 2003-2009 [14]

Not only efficiency but also tactics was analyzed in many papers , some example follows in the next tables .
Table 3 illustrates the types of tactics used in the 2010 Japan championship, where direct attack tactics came first by 66.6%, followed by the counter-attack with 33.3%, then combination attack with 8.3%.

| No. | Tactics types | Repetition | Percentages |
|---|---|---|---|
| 1 | Direct attacking | 8 | 66.6% |
| 2 | Combination attacking | 1 | 8.3% |
| 3 | Counter-attack | 4 | 33.3% |

Table 3: Judo tactics types in Japan 2010 championship [**15**]

The table 4,5 show the most performed throwing techniques, by japanese athletes during the world championship held in Japan in 2010, and the motor skill order and percentage. Instead table 6 shows the mutual relationship of technics for female world championships.

| Position (K) | K1 | K2 | K3 | Dominant techniques | The techniques most often performed | Number of attacks |
|---|---|---|---|---|---|---|
| 1 | 1 | 1 | 1 | Uchimata | Uchimata | 124 |
| 2 | 1 | 3 | 3 | Osoto gari | Seoi nage | 96 |
| 3 | 3 | 2 | 2 | Seoi nage | Deashi barai | 85 |
| 4 | 4 | 4 | 4 | Kouchi gari | Kouchi gari | 68 |
| 5 | 4 | 5 | 5 | Harai goshi | Ouchi gari | 57 |
| 6 | 6 | 6 | 6 | Kesa gatame | Tai otoshi | 53 |
| 7 | 6 | 8 | 7 | Kami shiho gatame | Osoto gari | 41 |
| 7 | 6 | 8 | 7 | Ashi guruma | Kosoto gari | 33 |
| 9 | 9 | 8 | 9 | Sasae tsurikomi ashi | Soto makikomi | 32 |
| 10 | 9 | 8 | 10 | Sode tsurikomi goshi | Harai goshi | 32 |

Tab4 techniques performed by japanese athletes at 2010 world championship [**15**]

| Skill type | Repetition | Percentages |
|---|---|---|
| Nage waza | | |
| Te waza | 2 | 13.3% |
| Koshi waza | 1 | 6.6% |
| Ashi waza | 5 | 33.3% |
| Sutemi waza | 5 | 33.3% |
| *Katame waza* | | |
| Osaekomi waza | 1 | 6.6% |
| Shime waza | 1 | 6.6% |

Tab.5 Motor skill percentage order of Japan 2010 world championships [**16**]

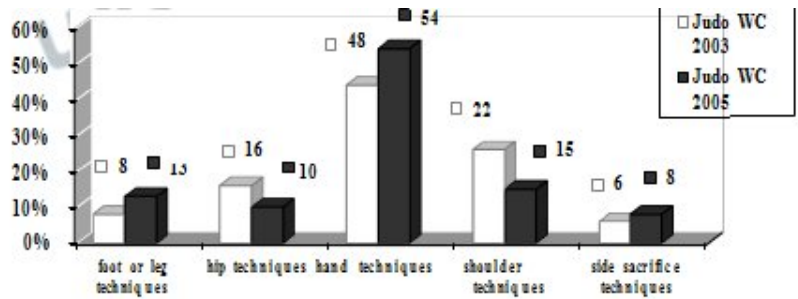

Tab.6 Mutual relation of the competition action model characterizations

in the middle weight category [17]

In the following diagrams are shown two other different ways to analyze in scientific manner high level Judo Competitions , it is possible to find the percentage statistics of throwing techniques applied in London Olympic 2012 , or the tactic study of attack directions preferred into a specific Athletes Couple System

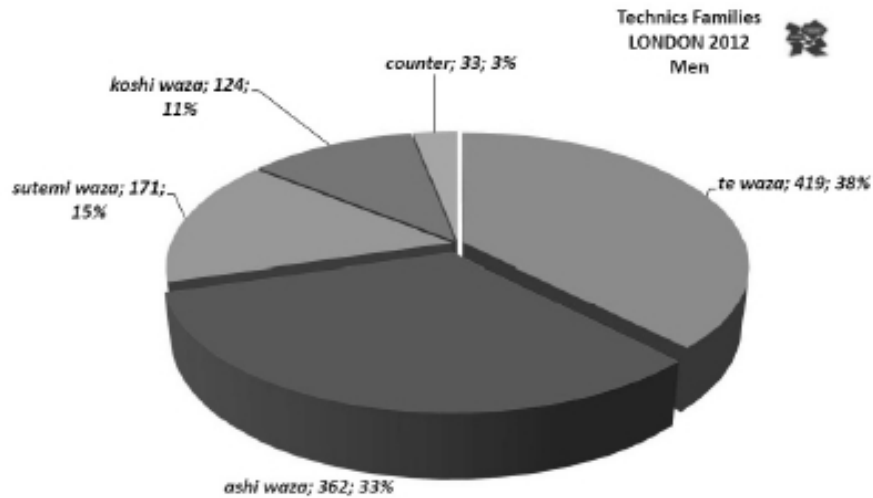

Diag.2 percentage of throwing in London 2012 [18]

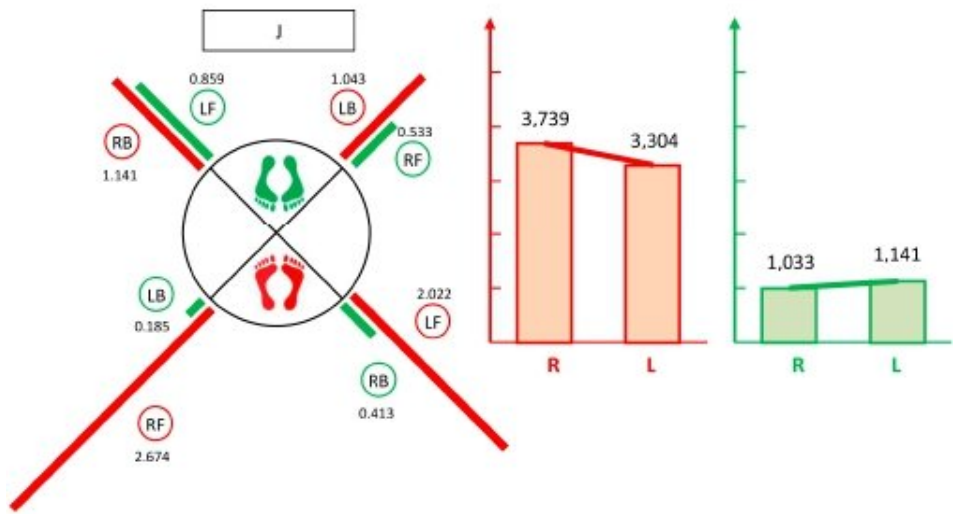

Diag.3 the attack direction preferred by a couple of judoka [19]

| Dominant techniques of Russian representatives | | | | | Techniques most often performed by Russian representatives | | |
|---|---|---|---|---|---|---|---|
| k | k1 | k2 | k3 | name of the technique | place | name of the technique | number of attacks |
| 1 | 1 | 2 | 1 | kuzure kesa gatame | 1 | kouchi gari | 81 |
| 1 | 1 | 2 | 1 | ude hishigi juji gatame | 2 | deashi harai | 53 |
| 3 | 3 | 2 | 3 | harai goshi | 3 | tai otoshi | 39 |
| 3 | 3 | 2 | 3 | kosoto gari | 4 | kosoto gari | 37 |
| 5 | 9 | 2 | 6 | tani otoshi | 5 | ouchi gari | 33 |
| 6 | 11 | 1 | 5 | uchimata | 6 | uchimata | 21 |
| 7 | 5 | 9 | 7 | seoi nage | 7 | seoi nage | 17 |
| 7 | 5 | 9 | 7 | tai otoshi | 8 | sasae tsurikomi ashi | 16 |
| 7 | 5 | 9 | 7 | okuri eri jime | 9 | sode tsurikomi goshi | 14 |
| 7 | 5 | 9 | 7 | uchimata sukashi | 9 | kosoto gake | 14 |
| 11 | 12 | 2 | 7 | sode tsurikomi goshi | 11 | tomoe nage | 13 |
| 11 | 12 | 2 | 7 | kosoto gake | 12 | sukui nage | 12 |
| 13 | 10 | 9 | 13 | deashi harai | 13 | ude hishigi juji gatame | 10 |
| 14 | 14 | 9 | 14 | sumi gaeshi | 14 | okuri eri jime | 7 |
| 14 | 14 | 9 | 14 | uki waza | 15 | osoto gari | 6 |

Tab. 7. Dominant techniques and the most often performed ones by Russian representatives during the OL 2012 in London [20]

| Variable | RF | RB | LF | LB | RFo | RBo | LFo | LBo |
|---|---|---|---|---|---|---|---|---|
| RF | 1.0000 | -.2774 | -.2336 | .0111 | .1086 | .2659 | .3241 | -.4281 |
|  | p = --- | p = .383 | p =.465 | p =.973 | p =.737 | p =.403 | p =.304 | p =.165 |
| RB | -.2774 | 1.0000 | .6112 | -.1572 | -.0944 | -.4528 | .0089 | .3819 |
|  | p =.383 | p =--- | p =0.035 | p =.626 | p =.770 | p =.139 | p =.978 | p =.221 |
| LF | -.2336 | .6112 | 1.000 | -.4166 | .4324 | -.4266 | -.4962 | .3702 |
|  | p =.465 | p =.035 | p =--- | p =.178 | p =.160 | p =.167 | p =.101 | p =.201 |
| LB | .0111 | -.1572 | -.4166 | 1.0000 | -.3387 | .7598 | .6439 | -.3974 |
|  | p =.973 | p =.626 | p =.178 | p =--- | p =.281 | p =.004 | p =.024 | p =201 |
| RFo | .1086 | -.0944 | .4324 | -.3387 | 1.0000 | .1236 | -.1582 | .0961 |
|  | p =.737 | p =.770 | p =.160 | p =.281 | p =--- | p =.702 | p =.623 | p =.766 |
| RBo | .2659 | -.4528 | -.4266 | .7598 | .1236 | 1.0000 | .5424 | -.5679 |
|  | p =.403 | p =.139 | p =.167 | p =.004 | p =.702 | p =--- | p =.068 | p =.054 |
| LFo | .3241 | .0089 | -.4962 | .6439 | -.1582 | .5425 | 1.0000 | -.3987 |
|  | p =.304 | p =.978 | p =.101 | p =.024 | p =.623 | p =.068 | p =--- | p =.199 |
| LBo | -.4281 | .3819 | .3702 | -.3974 | .0961 | -.5679 | -.3987 | 1.0000 |
|  | p =.165 | p =.221 | p =.236 | p =.201 | p =.766 | p =.054 | p =.199 | p =--- |

*(RF – right forward; RB – right back; LF – left forward; LB – left back; RFo – right forward opponents; RBo – right back*

Tab.8 Efficiency of throws with direction during OL 2008, the WC 2009-2011 and OL 2012 [20]

In these papers the attack ways that appear like the Japanese initiative studies, are not focalized on time attack or scope of attack as in the Japanese classification of Initiative, but more pragmatically by the tools or way of attack to obtain victory or indexes useful to evaluate effectiveness of each techniques., or study of directions connected with the attack way, but however for coaches is very difficult to single out some interesting information for high level preparation.

Most of the practical indication useful for coaches derive from Match Analysis by the classification tool.

More often coaches classify some basic or specific position seeing at the Couple of Athletes system, like: specific grips, relative body position, relative feet position, etc, and look for finding the best solution to solve in effective way such paradigmatic situations.

In the research field the current worldwide utilized classification of attack followed by researchers is grounded on three specific tools:

Direct Attack,
Combination
and Action Reaction.

This current point of view will be analyzed in the last part of the paper, with some useful addenda for coaching, this means that we will try to find some general rules that would be the basis to develop combinations and action reaction attack with a special attention to enhance such tactical tricks often applied in high level competitions.

## VI.1 *Direct Attack*

As the theoretical definition, direct attack is described in the following way: Tori launches a direct attack without preparation other than Tsukuri- kuzushi, all his energy in action, and body mass, stretched towards achieving it, sometimes direct attack is performed with only one grip or exceptionally without any stabilized grip.

However today the strength preparation and the skill of high level athletes are so increased that the success of this simple tool is really hard unless there is a big technical difference between the two adversaries.

From the practical point of view by the evaluation of classified match analysis positions, during time, coaches and athletes found very interesting ways to enhance such kind of attack like :       side way attack, to apply a Couple to Lever techniques, to change Couple techniques in rotational Lever   application, to utilize makikomi with both class of throwing, to apply transverse rotation after a  sagittal Couple throws, or to enhance Lever techniques by application of a Couple, etc.

 This type of attack  and its enhancement were deeply analyzed in a previous paper [ 8]

## VI.2 *Combinations*

Because it is not possible to consider and solve the infinitive match analysis situation that normally high level teams in different countries analyze in deep, we shrink our study starting not from many particular situations but from a more general definition which allow us a more flexible approach. We define combinations a special class of attack's iniziative based on the multiple application of the same or different throwing techniques.

In biomechanical terms this is a linear combination of the two basic tools to throw : Couple or Lever. Like Couple+Couple; Couple + Lever; Lever+Lever+Couple; etc.

In plane language we can assert that in this class generally it possible to find two main areas, that we call as practical definition:  repeated attack, and combination.

Repeated attacks are all these attack actions in which Tori launches a direct attack , allowed to react Uke , who avoids (or blocks or applies a counter) to immediately re-attack with the same technique. This process sometime could be useful and can surprise an opponent , especially when he is tired  (end of contest or during golden score) , in fact in that situation, very often the opponent, wrongly, resumes its original position and a second attack can destabilize Uke, obtaining the full point.
Combinations are always repeated  attacks applied by Tori, but more often combination are result of deep home preparation in the field of tactics.
Tori undertakes a series of true actions, all with the same sincerity, concluding with one of them .
The study of combination is strictly connected with the better use of the symmetry breaking produced in the Uke standing posture.
His theoretical development would be as follows : Tori with a true attack must cause sincerity reaction, Uke reacts, or bends, or turns in one directions which makes it vulnerable to different attacks of Tori .
Tori must find instantly the vulnerability of Uke to make an attack obtaining a projection.
More often today in high level competitions athletes act only in two ways, they  use combination in the same direction or in the opposite direction, because they are the main studied in the world, but the breaking symmetry concept needs a whole changing of approach, because it is connected in more subtle way to infinitive directions that are all useful to take initiative and score a point.  .
Biomechanics let us to analyze the basic mechanical properties that must be followed to obtain the victory.
If we analyze in deep the basic mechanics of combinations at light of Couple of Athletes it easy to single out their intimate connection with Couple system shifting velocity.
But how is the main biomechanical parameter that allows us to build effective combinations?
In dynamic  situation when the shifting speed changes the only useful biomechanical parameter is **relative distance** between athletes ruled by arms elongation, in fact this parameter changes in function of the specific throw utilized to obtain more or less contact to the adversary body.
In biomechanics Couple of Athletes System can be modeled as a spring mass system [00] then changing the spring elongation, the distance between athletes changes.

Now it is possible to understand the close connection among **Competition Invariants, shifting velocity, type of throws ( biomechanical throwing tool applied)** and **connected attack directions.**
On this basis it is possible to group all throwing techniques in three classes:

1. Throws applied at short distance (chica ma waza) connected to still or very slow speed of Athletes System,

2. Throws applied at medium distance ( ma waza), normal speed,

3. Throws applied at long distance (to ma waza) from normal to high speed of Athletes system.

it is important remember that shifting velocity ranges between 0,2 till to 0,5 m/s and it is close connected to the grips positions that fall and are classified into a Competition Invariants group.

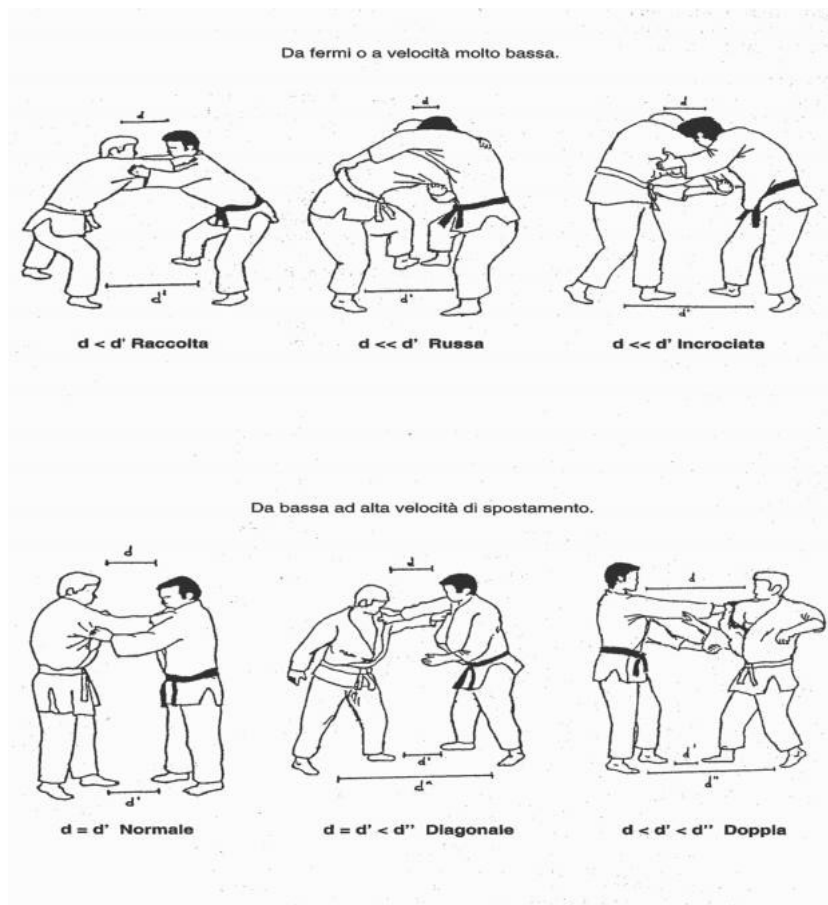

Fig 14 Competition Invariants group ( gripping position in competition)
Connected to shifting velocity and throws [21]

If Classical, Innovative and Chaotic throwing Techniques [22] are analyzed not only on the basis of the inner mechanics of throws but both at light of the best relative distance positioning to throw and at light of lengthening /shortening distance and possible shifting velocity.
It is possible to group all techniques in a new operative classification based on distance, this let us to find three specific groups ( Chica Ma Waza, Ma Waza and To Ma Waza).
These groups shown both in normal judo language and in biomechanical language the basic blocks that could be connected to build all possible combinations, without any specification about the directions or the applicative angles.

## Combinative Classification of Throws (*in function of attack distance*)

### Chica Ma Waza

(tight body contact – applied by body rotation at zero/ low velocity and strong grips-)

| | |
|---|---|
| • Uki Goshi | • Lever |
| O Goshi | • |
| Seoi Nage | • |
| Koshi Guruma | • |
| Kata Guruma | • |
| O Guruma | • |
| Tsurikomi Goshi | • |
| Tsuri Goshi | • |
| All Innovative henka | • |
| All Chaotic throws | • |
| • Hane Goshi | • Couple |
| Harai Goshi | • |
| O Soto Gari sideway | • |
| Uchi Mata | • |
| All Innovative henka | • |
| Very few Chaotic throws | • |

## Ma Waza
( medium distance applied with classical or double central grips)

| | |
|---|---|
| • Osoto Gari | Couple |
| Ouchi Gari | |
| All Innovative Henka | |
| • Sasae Tsurikomi Ashi | Lever |
| Harai Tsurikomi Ashi | |
| Uki Otoshi | |
| Sumi Otoshi | |
| O Soto Guruma | |
| Hiza Guruma | |
| Ashi Guruma | |
| Tai Otoshi | |
| All Innovative Henka | |
| All Chaotic Throws | |

## To Ma Waza
(applied at first contact, some are possible also with one sleeve grips only or theoretically without grip, all are Couple group techniques)

| | |
|---|---|
| De Ashi harai/barai | Couple |
| Ko Soto gari/gake | |
| Ko Uchi gari/barai | |
| Okuri Ashi harai/barai | |
| Uchi Mata with one grip | |
| O Soto Gari sideway Jumping | |
| All Innovative Henka | |

## Judo Combinations: a Biomechanical Principle

**A. To Change inter-athletes distance/contact both switching technique and final direction applying a linear combination of the two tools ( Couple and Lever)**

## *Four Application Ways*

1. Generally speaking combinations for athletes specialized in To Ma Waza ( long distance throws) that need timing, are organized on the basis of changing distance, from long one to shorter one with the support of right change of direction.
2. For athletes specialized in Chica Ma Waza (short distance throws) combination are grounded on changing direction: left/ right and vice versa or forward ,backward , or forward, forward, or backward, backward. Right/ side , and backward /side changing are useful and effective but less frequent also in high level competitions.
3. Athletes preferring Ma Waza (medium distance throws ) can more freely applying both change direction and shortening distance. Obviously sutemi are closing combinations throws, in whatever directions.
4. Balance is essential in every combination and direction changing, both for Ma Waza and Chica Ma Waza, more often Tori and Uke are not well balanced on the first attack but balanced as Couple System by their dynamic balance; Tori must fix with the unbalance action of his attack Uke body's position both on one or two legs in order to be able to change attack direction into the weak defensive side of Uke; this is possible when the unbalanced Uke, reacting to resist at a strong attack, becomes rigid and still.

Biomechanical analysis helps to systematize combination connecting the two tools to throws to the shifting velocity of Athletes Couple System.

V=0

Couple techniques    ( To Ma, Ma and Chica Ma)  Waza    <    Couple Lever

Lever techniques    ( Ma and Chica Ma)         Waza    <    Couple Lever

V= low/medium

Couple techniques    (To Ma, Ma and Chica Ma)  Waza    <    Couple Lever

It is not possible apply Lever techniques at low/medium speed Tori needs to stop Uke, and then it come back to the first situation

V= high

Couple techniques    ( To Ma Waza)             <    Couple Lever

It is not possible apply Lever techniques at high speed Tori needs to stop Uke, and then it come back to the first situation.

It is also possible, though difficult, to combine multiple techniques (three or more) to bring your opponent into a non defendable position. The better way to enhance such kind of linear combinations is to study how to connect throws that apply different biomechanical tool ( couple and lever) with the same leg; important to note that these connection are feasible only if the Couple of Athletes System is fixed.

 Translating in plain terms , this means to organize combinations like: Uchi Mata into Ko Uchi Gari into Tai Otoshi, or O Uchi Gari into Uchi Mata into Sumi Otoshi etc.

## VI.2.1  Combination Examples from high level competitions

The Biomechanical theory of Combinations is grounded on a practical point of view, as the pragmatic western vision of Initiative, today for one high performance coach is very important to give sound and easy understandable information to athletes improving their fighting capability.
In the following figures there are few examples of combinations, explained on the basis of the previous principle and with the utilization of the previous technical methodological classification reporting for clear connection the Japanese names of throwing techniques., the utilization of terms like shortening or lengthening distance must be understood  as change in contact athletes' bodies,  switching  different throwing techniques classified in function of distance of attack.
All these combinations sequences are taken by  judo high competition during the last forty years.

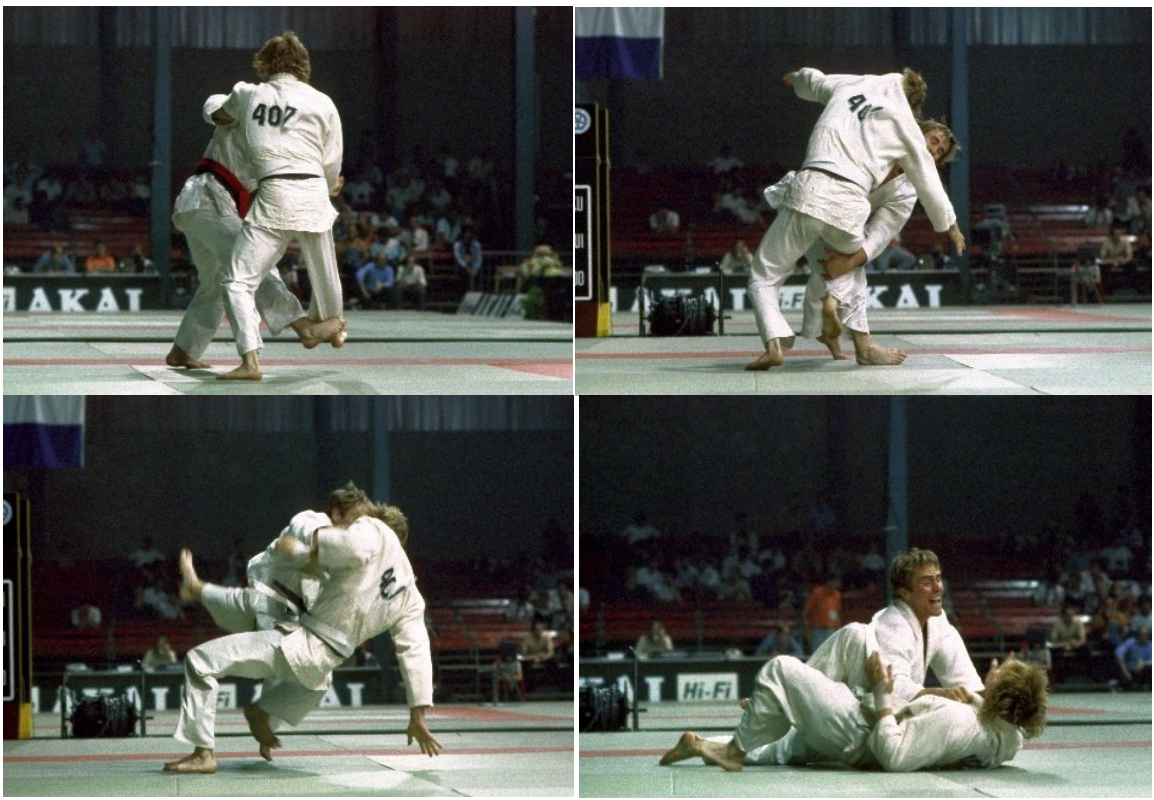

*Fig 15-18  Example of combination based on the shortening distance between athletes To Ma Waza  into Chica Ma Waza ( Ko Uchi Gari into Kuchiki Taoshi today not allowed  by referee regulations.)*  Adams against  Doherty (Finch)

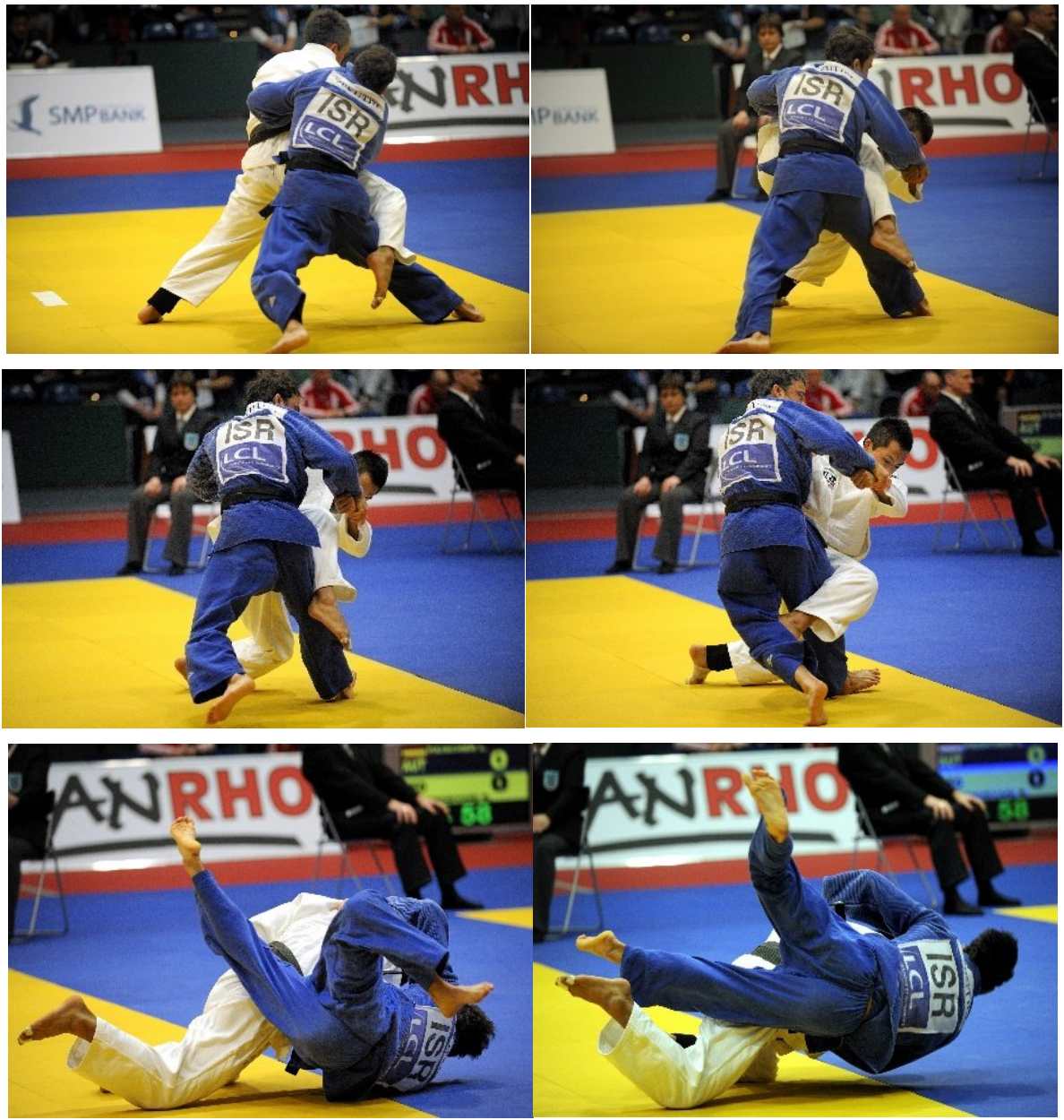

*Fig 19-24 Example of Chica Ma Waza into Ma Waza lengthening of distance with three changing of direction on a still and rigid Uke applying three Lever with same leg or three throws (O Soto Otoshi, sideway O soto Otoshi, sideway O Soto Guruma)*

Nomura against Yekutiel ( Finch)

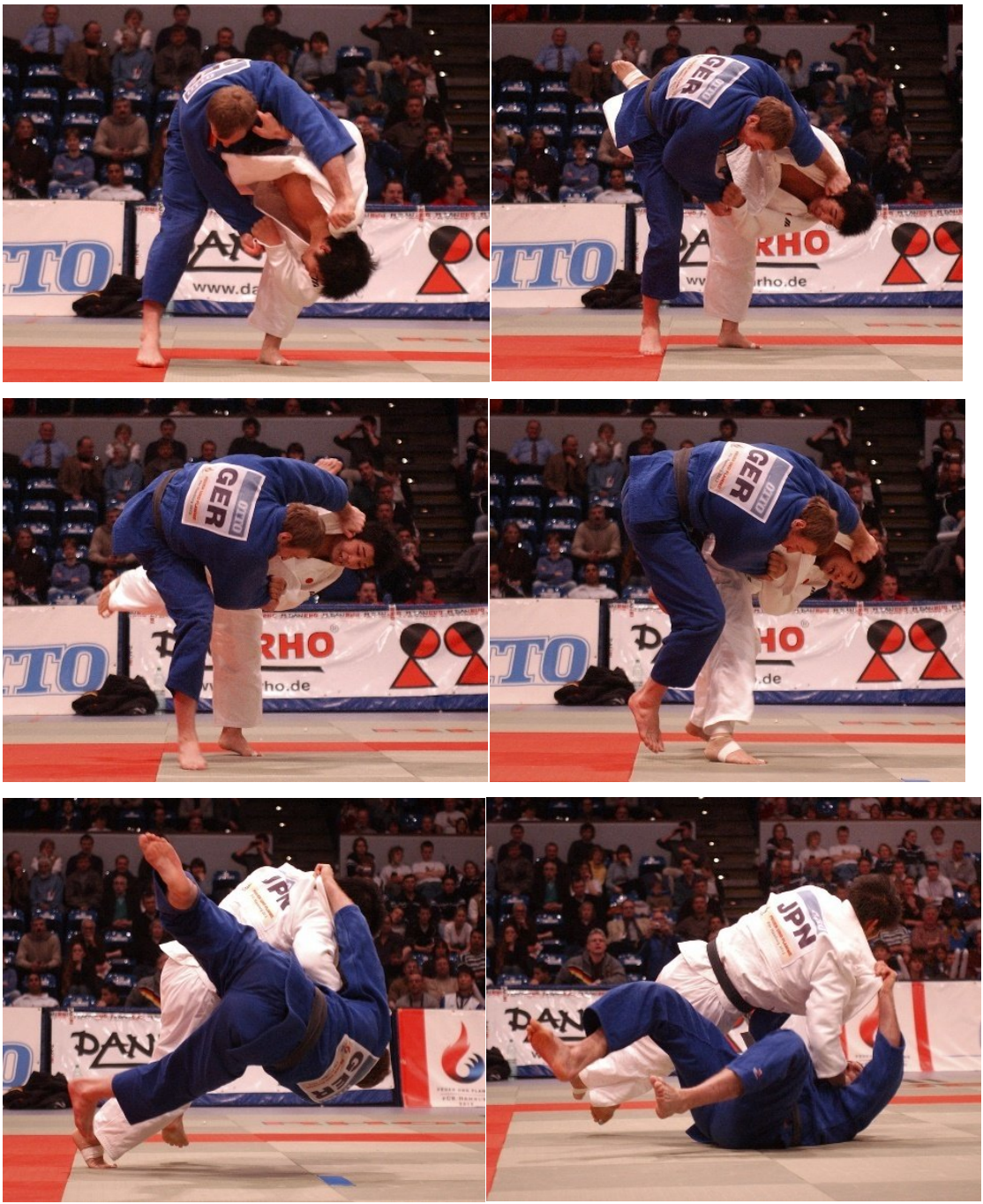

*Fig 25-30 Example of Chica Ma Waza into Ma Waza lengthening of distance with application of horizontal rotation and changing of Couple into Lever Tool.*
 *(Uchi Mata into Sumi Otoshi)*

Inue against Hubert ( Finch)

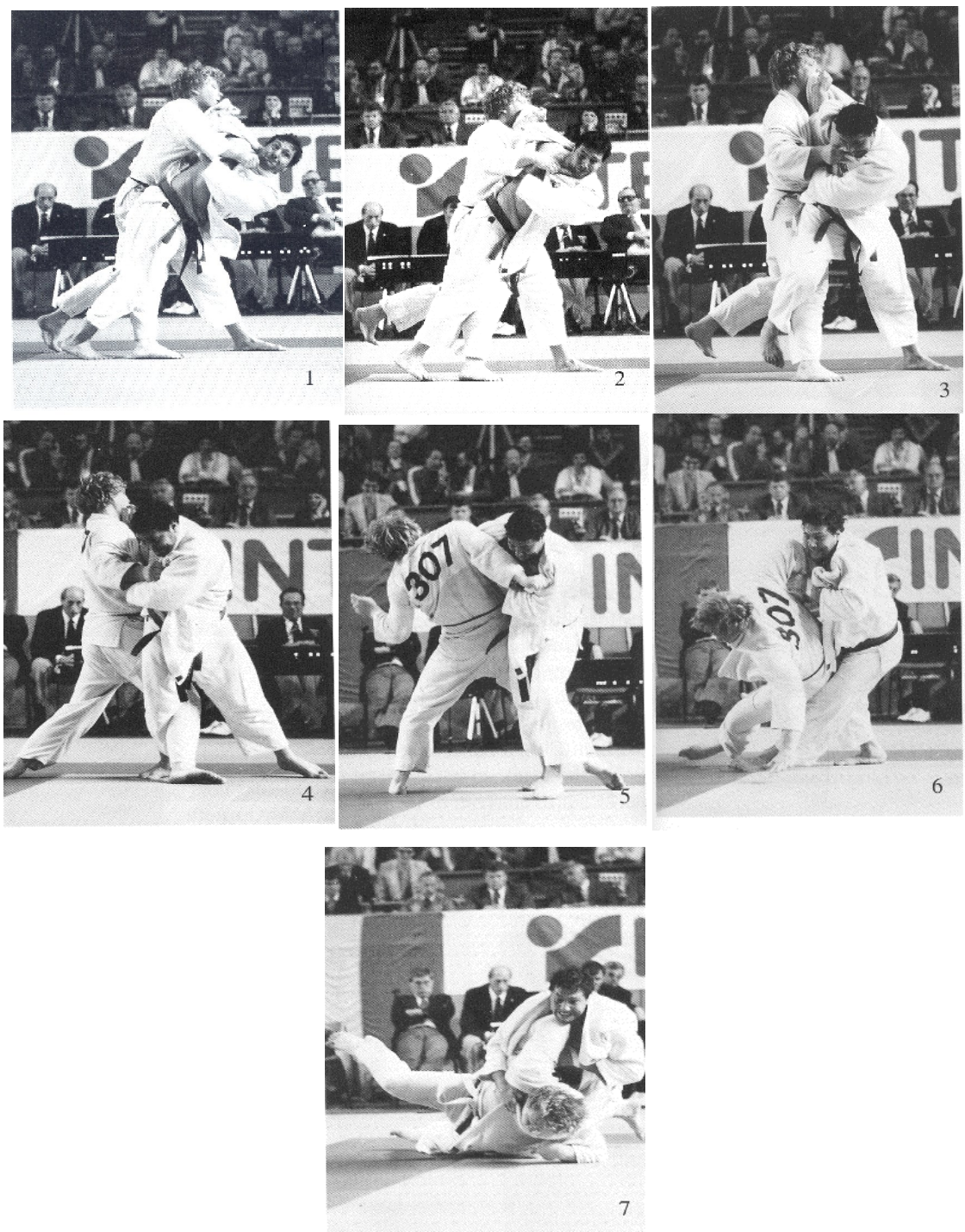

*Fig 31-37 Example of Chica Ma Waza into To Ma Waza lengthening of distance with changing of direction on a still rigid uke's leg, applying a combination of three Lever and one Couple or four different throwing techniques.*

*( Seoi into sideway O Soto Otoshi, O Soto Otoshi and Ko Soto Gake)*

Angelo Parisi. (Finch).

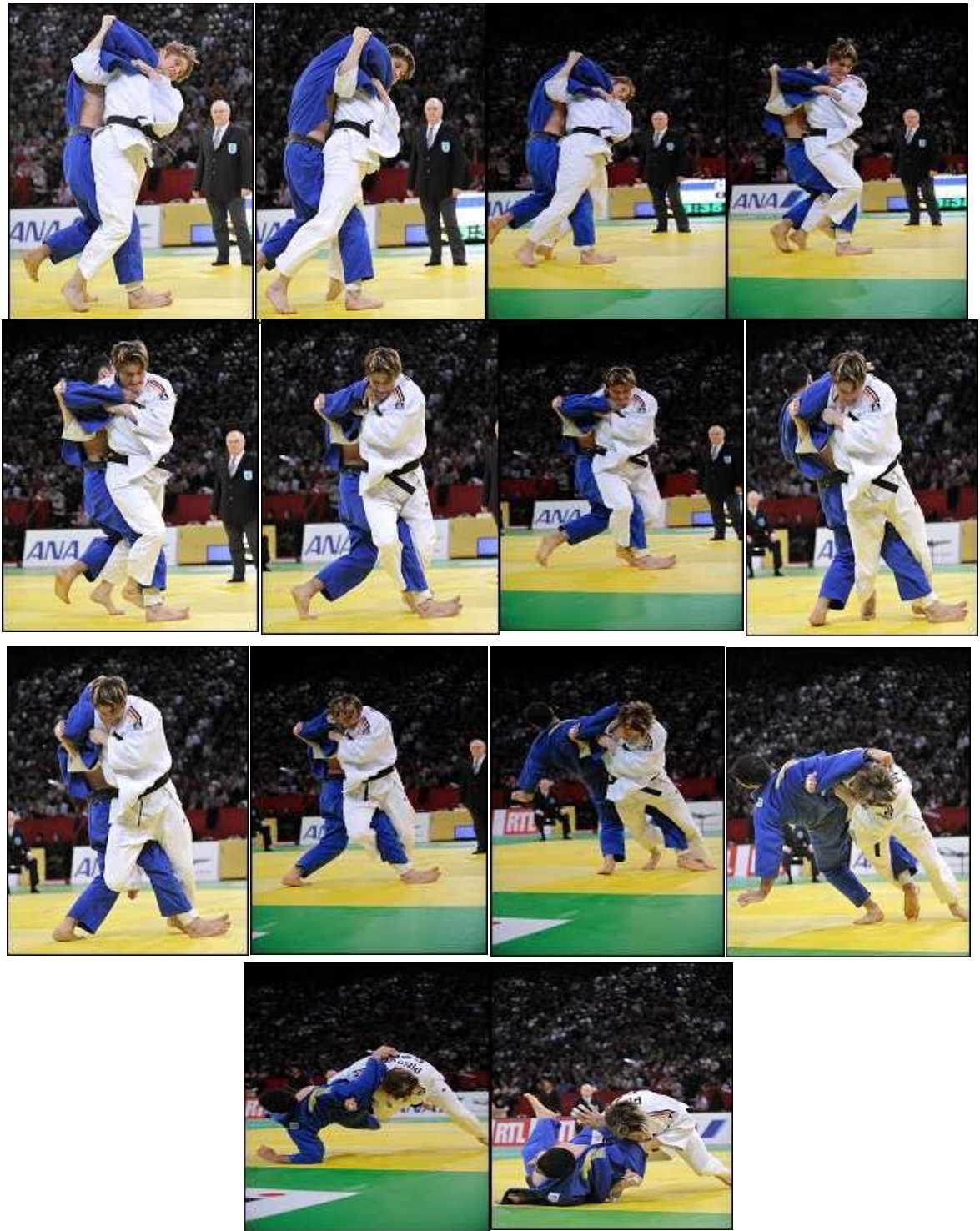

*Fig 38-51 Very similar to previous Example: Chica Ma Waza into To Ma Waza lengthening distance on a still rigid Uke, Tori applies a defensive blockage two Lever and one Couple with same leg, or three techniques (Seoi into O Soto Otoshi into O soto Gari)*
Pinske against Mammadov ( Finch)

Then biomechanical analysis of the previous situations let us to determine some interesting finding:

1. The only dynamical situation to build up complex combinations (2 or more techniques connecting Couple and Lever ) is at zero shifting velocity of athletes system.
2. In such specific dynamical situation, the fastest way to connect more techniques is to utilize the same one leg support position, changing goal at the moving "acting leg".
3. In term of "breaking symmetry" [7] Tori must break Uke's symmetry stopping his body in unstable equilibrium on one leg, increasing his stability on it, in order to totally block his mobility.
4. The previous position will be assured if Tori adds his own body weight on the stopped side of Uke's body.
5. Ten in such situation change direction to applied forces is only function of Tori's trunk rotation.
6. In theoretical way, combinations can be closed by every Sutemi, but as practical solution in high level competition that can't happen because too dangerous, in fact in such high dynamical situation it is possible to mistake something and undergo an hold ( Osaekomi)
7. The only Sutemi sometimes applied in competition is Tani Otoshi, because the mechanics of technique brings a final position more safe for Tori.
8. More frequently is utilized as final combination tool , both from female and heavy weight categories , the Makikomi trick connected with Lever techniques.

In general strategic terms, to take the initiative is advantageous also if no score is added at his own basket, in fact taking into consideration that currently analyzed athletes represent world's finest competitors in judo, it is very logical that their level of ducking attack (avoidance) is very high. Very often a large number of attacks does not always bring judoka direct points, but often leads to opponents' punishment due "passivity", and thus gaining an advantage. In addition, one should take into account that a large number of unsuccessful attempts as throwing opponent and contact to ground with a parts of the body that are not scored (Kinsa ) are both took  in account by referees in their evaluation in case of judgment, and filled as heavy psycho-physical pressure.

# VI.3 *Action Reaction*

*Theory*
When Judoka chooses to attack first, he can build an attack to cause an initial reaction of the opponent, he can exploit the utilized force by his opponent's action to its advantage.

*Pull/Push* : Judoka when attacks using a pull/push, this causes a reaction of the opponent.
So the judoka can transform its action creating a rear/forward unbalance.
The principle of action - reaction is the purest application of the Kano explication. "… if your adversary pushes you pull, and if he pulls you push…" .
Actually it could be based on the implementation of a movement trap to cause a natural adversary's reaction.
*Feint Principle* :Tori simulates an attack to react Uke in a direction that may operate to apply a decisive technique. The first attack, which is a preparation for the next attack may be more or less involved ; what matters is the apparent sincerity with which Tori must run for Uke gets caught lure.

*Practical application in high level competition*
In high level competitions today, it is not possible to apply, simply and directly the previous one methods.
Today athletes built a strong defensive system, and experience in high level fight furnishes them of a sound way to understand adversary's intention, then nor simple push/pull actions , nor simulated attacks can obtain the expected simple reaction, sometimes for example Uke reacts with not the expected reaction but with an unexpected ( prepared) avoidance movement in a more safe direction for him.
Then the only practical solution for Tori is to connect two well prepared and effective *real* attacks the second of which in the direction in which Uke may operate the only natural defensive action.
More often there are connected two couple techniques applied by the same leg, ore a Couple and Lever always applied by the same leg for saving attack time, making the action reaction trick more efficient and effective.

It follows four direction of action / reaction: forward / backward , backward /forward, left / right and right / left.

## VI.3.1 Action Reaction Examples from high level competitions

The Biomechanical theory of Action Reaction grounded on a practical point of view, as the pragmatic western vision of Initiative, it is simply connected to the application of two real attacks previously prepared in such way.
Tori applies a first real attack that can be stopped ( in the better way) in only one direction and connects the first one to an effective second attack in this specific only direction, often there are connected two Couple techniques applied by the same leg, or a Couple and Lever always applied by the same leg for saving attack time, making the action reaction more efficient and effective
Today for high performance coach is very important to prepare sound and easy understandable tactical tricks for the athletes to improve their fighting capability.
In the following figures there are two examples of action reaction tactical tricks, with the utilization of the previous technical methodology applied in high competition.

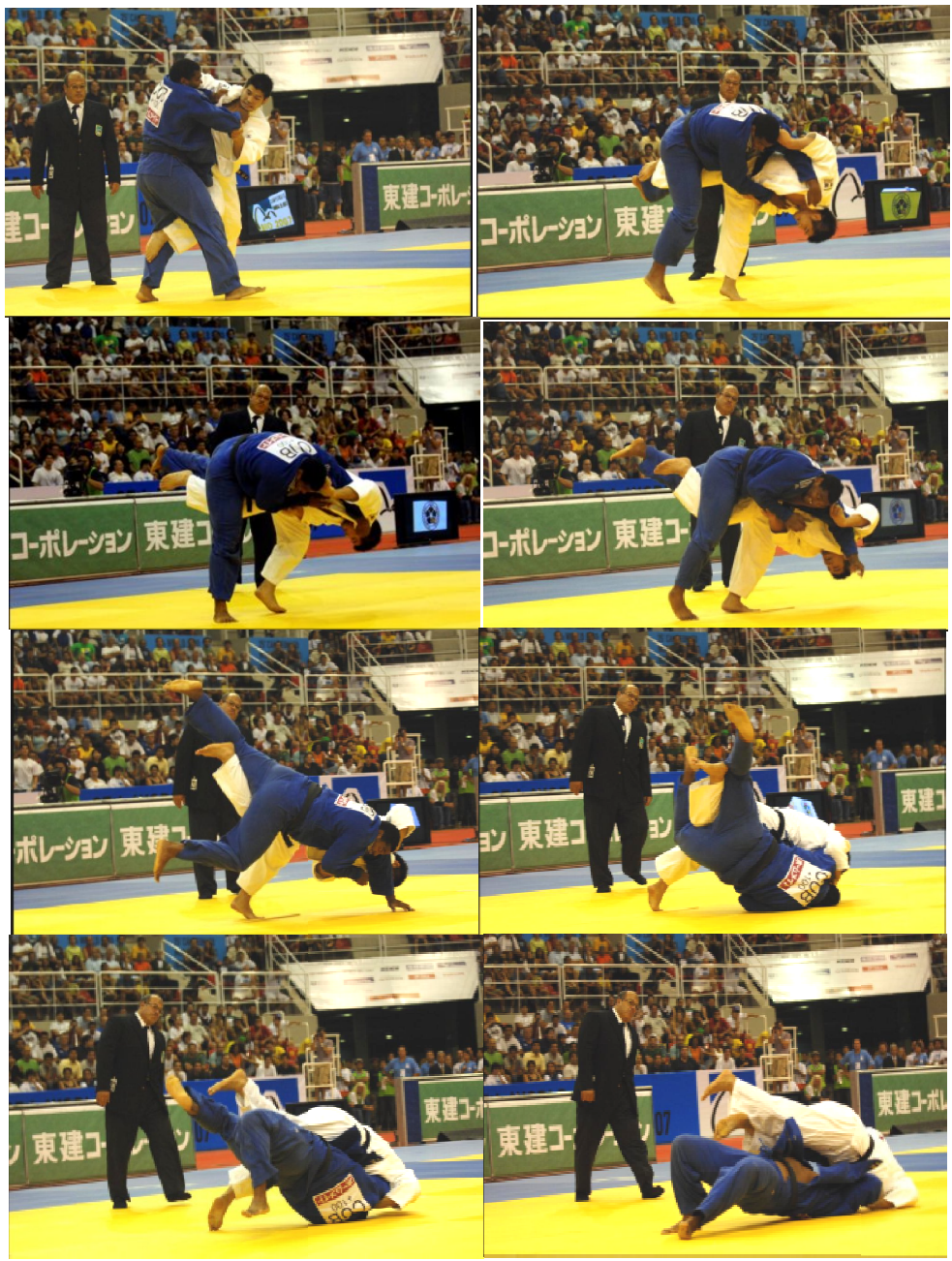

*Fig 52-59 Example of To Ma Waza into Ma Waza shortening of distance with application of backward/forward action-reaction on a natural simple reaction, using two Couple with the same leg. (O Uchi Gari into Uchi Mata)*

Inue against Bryson ( Finch)

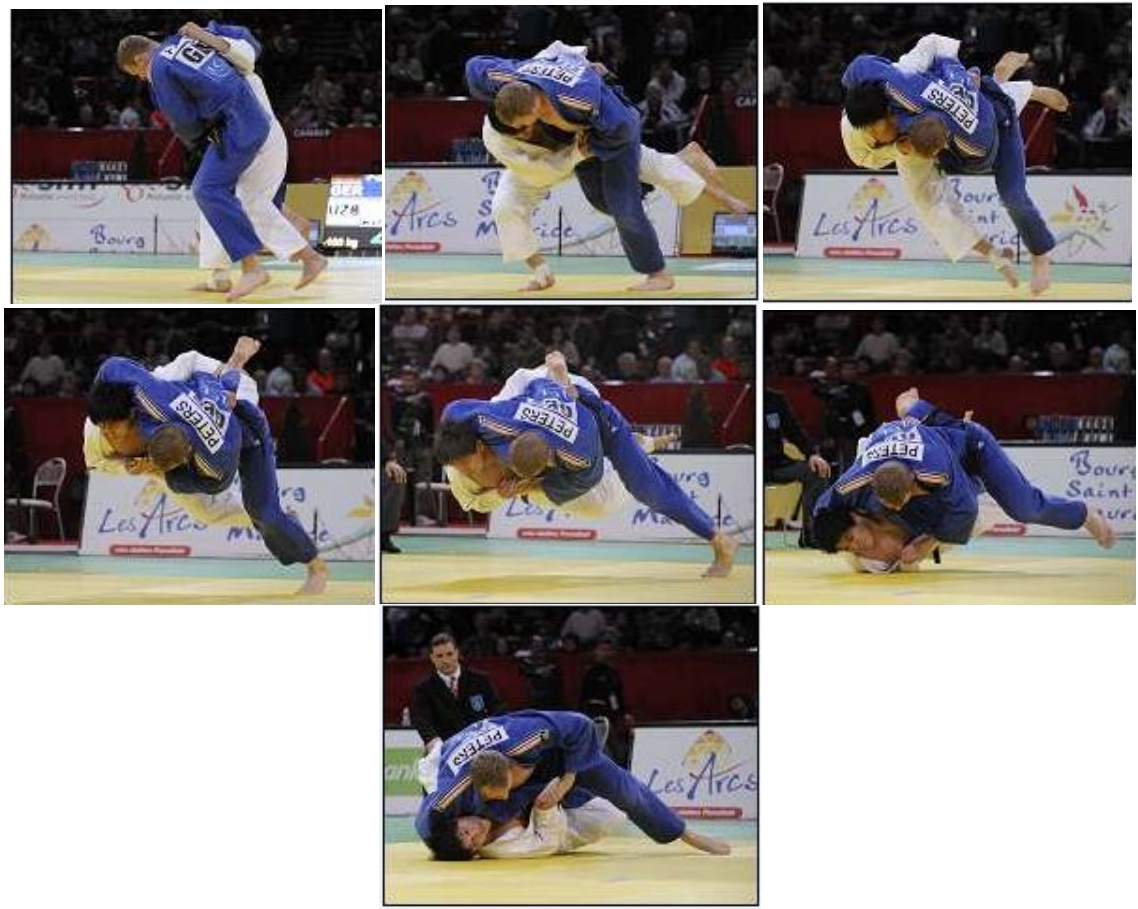

*Fig 60-66 Example of To Ma Waza into Ma Waza shortening of distance with application of backward/forward action-reaction on a natural simple reaction, using two Couple with the same leg. (rotational O Uchi Gari into Uchi Mata)*

Peters against Kubanov ( Finch)

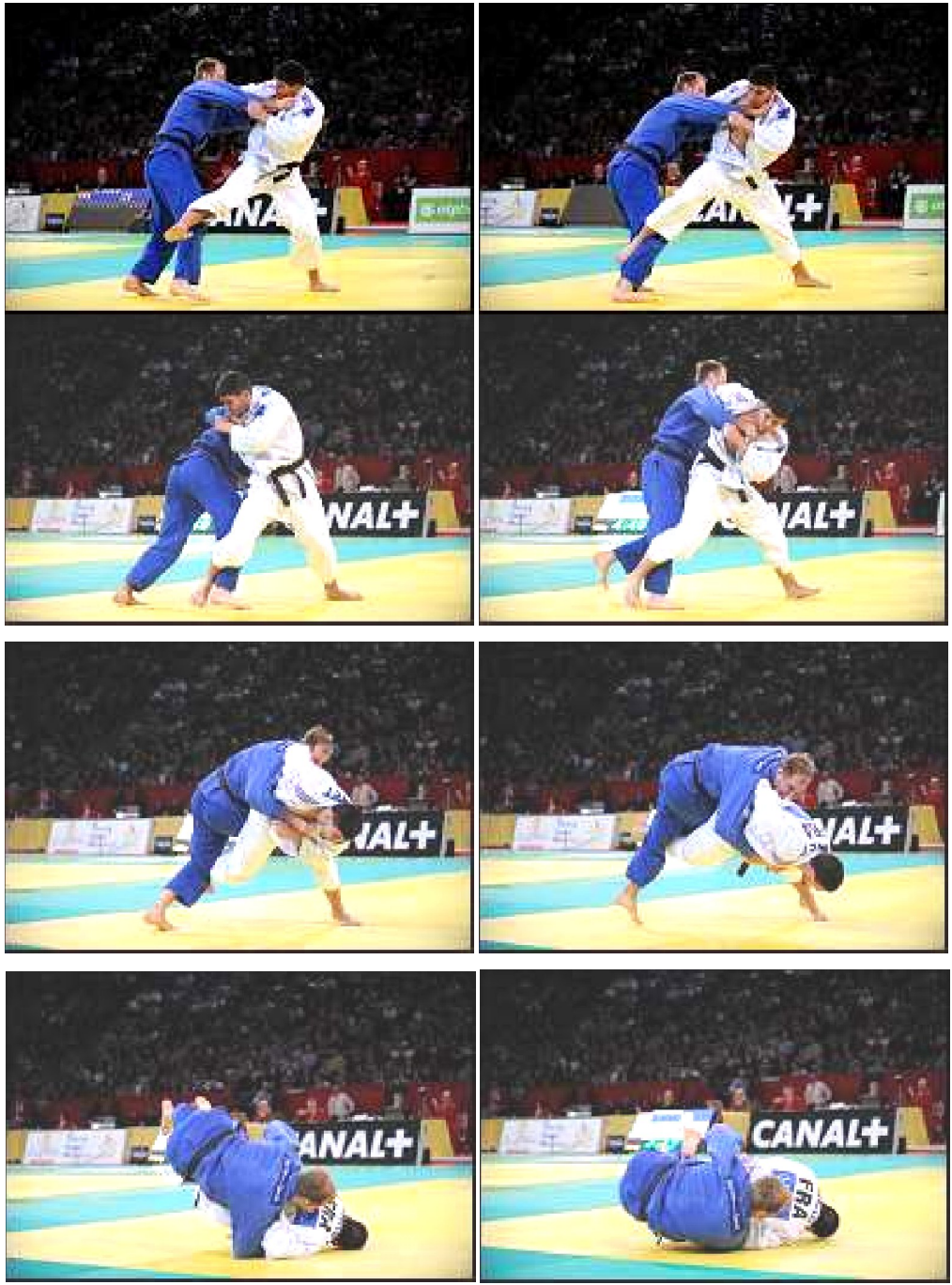

*Fig 67-74 Example of Ma Waza into Chica Ma Waza shortening of distance with application of backward/forward action-reaction on a more complex rotational avoidance, using two Couple with the same leg. (O Soto Gari into Uchi Mata)*

Maret against Van der Geest (Finch)

The biomechanical analysis of the previous high level competition situations let us to determine some interesting finding:

1. Normally action reaction attack is finalized at only two techniques connected by opposite attack directions.
2. Also for the action reaction attacks, the most effective situation is at zero shifting velocity of athletes system.
3. In such specific dynamical situation, the fastest way to connect two techniques is to utilize the same one leg support position, changing both direction of forces applied and goal at the moving "acting leg".
4. The breaking symmetry frequently applied for this actions is to bend or turn Uke's body stopped on his two feet.
5. The attack directions more often utilized are backward/ forward, forward/ backward, less frequent (but always possible) are left sideway/right sideway and vice versa.
6. In term of double central grip, the flexibility to apply double attack directly on the right and on the left by Ma Waza, is frequently connected with forward/ backward change of direction.

## VII Conclusions

There are many years that in the researches papers high level competition are analyzed in deep and tactics is grouped in three attack ways ; Direct attack, Combination, Action-reaction tricks.

Direct attack and his enhancements was analyzed extensively in a previous paper.

In this paper the biomechanical analysis of the remaining two tactics tricks utilized in high level competition: Combination and Action –reaction  let us to obtain very interesting findings for coaching .

It was able to single out the mechanics of combination, connected to one of the Competition Invariant class. Such connection linked to the shifting velocity of Couple of athletes system, let us to understand that combination are grounded on the changing in contact/distance of the athletes bodies by means of different group of throws classified as:  Chica ma waza, ma waza, to ma waza.

Furthermore  biomechanical analysis of the high level combination let us to determine some interesting finding:

1. The only dynamical situation to build up complex combinations (2 or more techniques connecting Couple and Lever ) is at zero shifting velocity of athletes system.
2. In such specific dynamical situation, the fastest way to connect more techniques is to utilize the same one leg support position, changing goal at the moving  "acting leg".
3. In term of "breaking symmetry" Tori must break Uke's symmetry stopping his body in unstable equilibrium on one leg, increasing his stability on it, in order to totally block his mobility.
4. The previous position will be assured if Tori adds his own body weight on the stopped side of Uke's body.
5. Ten in such situation change direction to applied forces is only function of  Tori's trunk rotation.
6. In theoretical way, combinations can be closed by every Sutemi, but as practical solution in high level competition that can't happen, because too dangerous, in fact in such high dynamical situation it is possible to mistake something and  undergo an hold ( Osaekomi)
7. One Sutemi sometimes applied in combination during competition is Tani Otoshi, because the mechanics of technique brings a final position more safe for Tori.
8. More frequently is utilized as final combination tool , both from female and heavy weight categories, the Makikomi trick connected with Lever techniques

As regards to Action-Reaction tricks, due to the outstanding physical preparation and the increasing fighting experience of athletes it is very difficult to apply them based only to simple push/pull action.

Today this tricks must be integrated by a real technical attack often on a motionless Uke based on breaking symmetry on two feet.

The biomechanical analysis of the Action-Reaction situations let us to determine some interesting finding:

1. Normally action reaction attack is finalized at only two techniques connected by opposite attack directions.
2. Also for the action reaction attacks, the most effective situation is at zero shifting velocity of athletes system.
3. In such specific dynamical situation, the fastest way to connect two techniques is to utilize the same one leg support position, changing both direction of forces applied and  goal at the moving  "acting leg".

4. The breaking symmetry frequently applied for this actions is to bend or turn Uke's body stopped on his two feet.
5. The attack directions more often utilized are backward/ forward, forward/ backward, less frequent (but always possible) are left sideway/right sideway and vice versa.
6. In term of double central grip, the flexibility to apply double attack directly on the right and on the left by Ma Waza, is frequently connected with forward/ backward change of direction.